\documentclass[usenatbib]{mn2e}

\usepackage{graphicx}
\usepackage{lscape}
\usepackage{txfonts}
\usepackage{url}
\usepackage{lastpage}

\title[Satellite galaxies around present-day massive ellipticals]{Satellite
galaxies around present-day massive ellipticals}

\author[Pablo Ruiz, Ignacio Trujillo \& Esther M\'armol-Queralt\'o]{Pablo Ruiz$^{1}$ \thanks{E-mail:
ruihern@gmail}, Ignacio Trujillo$^{1,2}$ and Esther M\'armol-Queralt\'o$^{1,2,3}$\\
$^{1}$Departamento de Astrof\'{\i}sica, Universidad de La Laguna, E-38205, La 
Laguna, Tenerife, Spain\\
$^{2}$Instituto de Astrof\'{\i}sica de Canarias, c/ V\'{\i}a L\'actea s/n, 
E-38205, La Laguna, Tenerife, Spain\\
$^{3}$ Institute for Astronomy, University of Edinburgh, Royal Observatory, Edinburgh, EH9 3HJ}
\begin{document}

\date{}

\pagerange{\pageref{firstpage}--\pageref{lastpage}} \pubyear{2014}

\maketitle

\label{firstpage}

\begin{abstract}

Using the spectroscopic New York University Value-Added Galaxy Catalogue and the photometric
photo-z catalogues of the Sloan Digital Sky Survey Data Release 7, we have explored the satellite
distribution around $\sim$1000 massive
(M$_\star$$\gtrsim$2$\times$10$^{11}$M$_\odot$) visually classified elliptical galaxies  down
to a satellite mass ratio of 1:400 (i.e.
5$\times$$10^{8}$$\lesssim$M$_{sat}$$\lesssim$2$\times$10$^{11}$M$_\odot$). Our host galaxies
were selected to be representative of a mass complete sample. The satellites of these
galaxies were searched within a projected radial distance of 100 kpc to their hosts. We have
found that only 20-23 per cent of the massive ellipticals have at least a satellite down to a mass
ratio 1:10. This number increases to 45-52 per cent if we explore satellites down to 1:100 and is
$>$60-70 per cent if we go further down to 1:400. The average projected radial distance of the
satellites to their hosts for our whole sample down to 1:400 is $\sim$59 kpc (which can be
decreased at least down to 50 kpc if we account for incompleteness effects). The number of
satellites per galaxy host only increases very mildly at decreasing the satellite mass. The
fraction of mass which is contained in the satellites down to a mass ratio of 1:400 is 8 per cent of
the total mass contained by the hosts. Satellites with a mass ratio from 1:2 to 1:5 (with
$\sim$28 per cent of the total mass of the satellites) are the main contributor to the  total
satellite mass. If the satellites eventually infall into the host galaxies, the merger
channel will be largely dominated by satellites with a mass ratio down to 1:10 (as these
objects have 68 per cent of the total mass in satellites).

\end{abstract}


\section{Introduction}

A combination of major and minor merging has raised in the last few years as the most likely channel of
size and mass growth of present-day massive (M$_\star$$\gtrsim$10$^{11}$M$_{\sun}$) ellipticals. A
number of observational as well as theoretical studies support this mode of growth. According to that
picture, present-day most massive ellipticals created the bulk of their mass in a short but very
intense starburst event at z$\gtrsim$2
\citep[e.g.][]{Keres2005,Dekel2009,Oser2010,Ricciardelli2010,Wuyts2010,Bournaud2011}. The initial
structural configuration of these galaxies was very compact. Later on, a continuous bombardment process
by the satellites orbiting these objects produced the envelopes that we see today surrounding these
galaxies \citep{Khochfar2006, Oser2010,Feldmann2011}. There are several observations that fits well
within this scheme. To  name a few, we have for example evidences for a progressive size evolution of
the massive galaxies since z$\sim$3 \citep[e.g.][]{Trujillo2007,Buitrago2008} produced mainly by the
formation of the outer regions \citep[e.g.][]{Bezanson2009,Hopkins2009a,vanDokkum2010}. We also have 
evidences showing that the size evolution of the massive galaxies does not depend on the age of their
stellar populations \citep{Trujillo2011} nor in their intrinsic sizes \citep{Diaz2013}, suggesting a
growth mechanism  external to the galaxy properties. Moreover, the average velocity dispersion of the
massive galaxy population has decreased mildly since z$\sim$2, in good agreement with the theoretical
expectations based on galaxy merging \citep[e.g.][]{Cenarro2009}. All these results have made more
unlikely that the channel growth of massive galaxies could be driven by AGN or supernova feedback
\citep{Fan2008,Fan2010,Ragone2011}. Nonetheless, all the above observational evidences test the growth
of massive galaxies only indirectly. It is timely then to conduct a detailed analysis of the properties
of the satellites that promote the growth of the massive galaxy population.

Several works have studied in detail what are the properties of the satellites surrounding
the massive elliptical galaxies over cosmic time
\citep{Jackson2010,Nierenberg2011,Nierenberg2013,Man2012,Newman2012,Marmol-Queralto2012,Marmol-Queralto2013,Huertas2013}.
\cite{Marmol-Queralto2012} have found that the fraction of massive galaxies with satellites
of a given mass ratio (1:100 up to z=1 and 1:10 up to z=2) has remained constant with time.
This constancy of the number of satellites surrounding the massive galaxies is in good
agreement with semianalytical predictions based on the $\Lambda$ cold dark matter ($\Lambda$CDM) scenario
\citep{Quilis2012}. However, the theoretical estimates over-predict the fraction of massive
galaxies with satellites down to 1:100 mass ratio by a factor of $\sim$2. It is unclear
though how relevant could be the effects of incompleteness in the observational studies. 

The goal of this paper is to create a local (z$\sim$0) reference that can be used to anchor the
evolution of the satellite population of massive galaxies at higher redshifts. In particular, we
concentrate on this paper on present-day massive ellipticals which are the galaxies which show the
largest size evolution with cosmic time. To facilitate the comparison with high-z studies, the way this
local reference of massive elliptical galaxies is created is based only on the visual morphology of the
galaxies. We do not make any color selection of our galaxies to avoid biasing our sample towards either
young or old galaxies. In addition, we do not make any previous selection of our sample based on
environmental density criteria. To reach our aim, we have used the morphological galaxy catalogue of
\cite{Nair2010}.  Because of the vicinity of our objects (z$<$0.1), we can probe the satellite distribution
of these massive ellipticals down to a satellite stellar mass M$_{sat}$$\sim$5$\times$10$^{8}$M$_{\sun}$. Our
work also includes another necessary exercise. We have explored the differences between the satellite
populations when using a purely spectroscopic redshift sample or a photometric redshift one. This
comparison is worth doing as in higher redshift samples the ability of selecting satellites using
spectroscopic data alone is severely compromised by the faint apparent magnitudes of these objects. 

Although it is not our primary goal, the large number of massive galaxies studied in this paper could be
used in future works to make a direct test of the $\Lambda$CDM predictions about the number of
satellites surrounding the most massive galaxies in the present-day Universe \citep[see for
instance][]{Liu2011,Wang2012}. Finally, our study will allow us to explore which is the most likely merging channel of
present-day massive ellipticals. We will quantify which type of satellites will contribute most to the
mass increase of their host galaxies in case they eventually merge with the main object.  Undoubtedly,
this local study, together with other works at higher z \citep[see e.g.][]{Ferreras2014}, will allow us to explore whether the merging
channel (i.e. which type of satellite is the most likely to contribute to the mass and size growth)
has changed with time or not.

The paper is structured as follows. In Section ~\ref{sec:data}, we describe our sample of massive and
satellite galaxies. Section  ~\ref{sec:selection} explains the satellite selection criteria and the
methods used to clean our sample from background and clustering effects. Our main results are presented
in Section \ref{results}. The distribution of the mass contained in satellites surrounding the massive
ellipticals is shown in Section \ref{sub:merging channel}. Section \ref{sec:discussion} discusses the
main results of this paper and finally our work is summarized in Section~\ref{sec:conclusions}. Hereafter,
we assume a cosmology with $\Omega_{\rm m}= 0.3$, $\Omega_{\rm \Lambda}= 0.7$ and H$_0 = 70$
km~s$^{-1}$~Mpc$^{-1}$. 


\section{The data}\label{sec:data} 

Our study is based on two different datasets: the sets of massive elliptical (host) galaxies and the
samples of satellites around them. On what follows, we describe how these samples were obtained.

\subsection{Samples of massive elliptical galaxies}\label{sec:smeg}

Our samples of massive elliptical galaxies have been obtained from the morphological catalogue
published by \citet{Nair2010} (hereafter NA10). This catalogue comprises 14034 galaxies with detailed
visual classification and available spectra from the Sloan Digital Sky Survey (SDSS) DR4
\citep{Adelman-McCarthy2006} in the spectroscopic redshift range 0.01$<$z$<$0.1 down to an apparent
extinction-corrected limit of g$<$16 mag. Within this catalogue, we select the elliptical galaxies
(i.e. c0, E0 and E+; T-Type class $-5$), obtaining a sample of 2723 visually classified elliptical
galaxies. In order to have the more up to date spectroscopic redshift determination for these galaxies,
we cross correlated that catalogue with the spectroscopic NewYork University Value-Added
Galaxy Catalogue (NYU VAGC;~\citet{Blanton2005}) based on the
SDSS Data Release 7 (DR7;~\citet{Abazajian2009}), obtaining 2654 galaxies
common in both data sets.   

In addition to a visual classification of the galaxies, the catalogue from NA10 provides other
important parameters as  their stellar mass  \citep[based on][]{Kauffmann2003}. For consistency with
the catalogue of satellite galaxies where we have estimated the stellar masses using the
\citet{Bell2003} recipe, we have also measured the stellar mass of our host elliptical galaxies using
the same technique. This is done as follows. We take the SDSS bands magnitudes of our objects corrected
from Galactic extinction \citep{Schlegel1998}. Later on, we apply the K-correction provided by the NYU
VAGC \citep[][]{Blanton2007} both in the r band and in the g-r colour in order to have these two
magnitudes measured in the rest-frame of our objects. Then, using  the prescription of \citet{Bell2003}
we estimate the mass-to-luminosity (M/L) ratio as a function of  the rest-frame colour, assuming a \citet{Kroupa2001} initial
mass function (IMF)\footnote{We have selected this IMF to facilitate the comparison of our results with
previous results conducted at higher redshift \citep[e.g.][]{Marmol-Queralto2012} and with numerical 
simulations \citep[e.g.][]{Quilis2012}.}. Thus, the  M/L ratio in the r band is
estimated like this:

\begin{equation}\label{formula_mass_lum}
log(M/L)_{\rm r} = a_{\rm r} - b_{\rm r} (g-r) -0.15,
\end{equation}

where $a_{\rm r}=-0.306$ and $b_{\rm r}=1.097$ are the specific coefficients applied to SDSS for determining
the M/L ratio in the r band and 0.15 is subtracted to get the results assuming a Kroupa IMF.

After computing  $(M/L)_{\rm r}$, we can directly estimate the stellar mass  using the next
relation:

\begin{equation}\label{formula_mass}
log(M/M_\odot) = log(M/L)_{\rm r} - 0.4(M_{\rm r} - M_{\rm _\odot,r}),
\end{equation}

where $M_{\rm r}$ is the absolute magnitude of the galaxy and $M_{_\odot,\rm r}$=4.68 the
absolute magnitude of the Sun in the SDSS r band. We have compared our stellar mass
estimates with the measurements provided by NA10. We obtain the following results: our
stellar masses derived with   \citet{Bell2003} are above (0.09 dex) than those derived  by
NA10 which are based on \citet{Kauffmann2003}. This is not surprising taking into account the
different methodologies and stellar population models used in both estimates of the stellar
mass. The mean difference between both stellar mass determination has a scatter of 0.1 dex.
We consider this value as our typical uncertainty at estimating the stellar mass of our host
galaxies.

\begin{figure}
\centering
\includegraphics[width=1.0\columnwidth,clip=true]{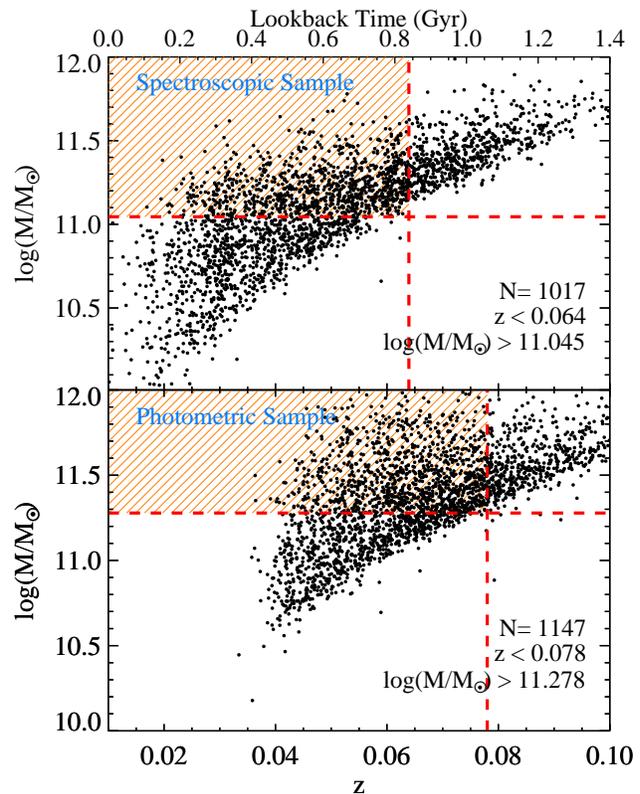}

\caption{Stellar mass versus redshift for the massive elliptical (host) galaxies of our spectroscopic
(upper panel) and photometric (lower panel) samples.  The vertical and horizontal red dashed lines
establish the maximum redshift   and minimum mass limits used in this paper to maximize the number of
massive elliptical galaxies within a complete mass subsample.}

\label{sample}
\end{figure}

Once we have the stellar masses as well as the spectroscopic redshifts of our host galaxies, we build a
mass complete subsample of host galaxies within a given redshift range. There is some observational
evidence suggesting that the satellite population could depend on the mass of their host galaxies
\citep[e.g.][]{Wang2012}. Consequently, if our host sample were not complete in mass, we will be mixing
hosts with different satellite populations along our redshift range of exploration, eventually biasing
our results. The building of this complete subsample of host galaxies is done as follows. In the
stellar mass - redshift plane (see Fig.~\ref{sample}; upper panel), we select which combination of
stellar mass and redshift maximizes the number of massive galaxies within those ranges. We find 1017 
massive ellipticals above 1.1$\times$10$^{11} M_\odot$ and z$<$0.064. This is the sample of host
galaxies that we will use in the rest of the paper when referring to the spectroscopic host sample.

We have repeated the above exercise but this time using photometric redshifts for the host galaxies. We
have conducted this exercise since we are  pushing the analysis of the satellite population down to
faint magnitudes where the redshift determination is only photometric. For this reason, for consistency,
is necessary to have also the redshifts and the stellar masses of the host galaxies determined using
photometric redshifts as well. It is worth noting that the sample of host galaxies built that way is
related with the spectroscopic sample but it does not necessary contain the same all galaxies. For
instance, the source of photometric redshifts we have used  (which we will describe
later) provides larger redshifts (z$\gtrsim$0.04) for the host galaxies than the ones measured spectroscopically. As the
photometric redshifts estimation has its own biases and uncertainties, our analysis using the
photometric sample is fully independent of the spectroscopic analysis. In fact, this analysis has to be
considered as an alternative analysis to the one using the spectroscopic sample. In other words, the
exercise conducted in this paper explores how different the results would be in case we were only
having spectroscopic or photometric redshifts. Nonetheless, throughout the paper we will often compare
both analysis to check the consistency of our results. The building of the mass complete subsample in
the case of the photometric sample is done as in the spectroscopic case. In the photometric stellar
mass - redshift plane (see Fig.~\ref{sample}; bottom panel), the sample is maximized (1147 objects) for
masses above  1.9$\times$10$^{11} M_\odot$ and z$<$0.078. The spectroscopic and photometric sample have
696 hosts in common (i.e. $\sim$65 per cent of the sample).

\subsection{Samples of satellite galaxies}

The samples of satellites surrounding our host galaxies are based on the following two criteria: their
stellar mass and their proximity (both in spatial projected distance as well as in redshift)  to our
host galaxies. We will describe in Section \ref{sec:selection} what are the exact criteria in distance
and redshift used to select our satellite candidates. In this section, we describe how the
redshift and stellar mass have been determined for the pull of galaxies that are used to select the
satellite candidates.

As the basis for the redshifts of our potential satellite galaxies we have used both the spectroscopic
NYU VAGC and photometric ('photo-z') redshift catalogues of the SDSS DR7. The NYU VAGC based on the SDSS
DR7 spectroscopic data base contains over 900000 spectroscopically confirmed galaxies and it is roughly
complete down to r$\sim$17.7 mag.  Similarly, our
photometric catalogue comprises the galactic 'primary' sources with photometrically estimated redshifts
and K-corrections from 'photo-z' data base. This is a large and continuous data set of galaxies within the
SDSS DR7 coverage and whose 95 per cent
completeness\footnote{http://www.sdss.org/dr7/products/general/completeness.html} is estimated to be
around r$\sim$21.5 mag. The photometric redshift uncertainty is 0.022.

As mentioned before, we have estimated the stellar masses of our satellite candidates using the
prescription given by \citet{Bell2003}. This prescription is based on the g-r colour. Consequently, we
need to account for the photometric errors both in the g and r bands in order to assure this colour is
measured with enough confidence to provide reliable stellar mass estimates. For this reason, in
addition to the magnitude limit in the r band we have used above, we also demand that the photometric
errors at estimating the number counts of each galaxy will be less than 3$\sigma$ the expected error at
measuring their number counts. In other words, acceptable photometric errors for each object are those
where error(counts)$\lesssim$3$\times$sqrt(counts+$\sigma_{sky}$$^2$). $\sigma_{sky}$ is the
uncertainty (in counts) at measuring the sky value in each band\footnote{Typical values for the sky in
the SDSS images are 24.88 counts (g-band) and 23.96 counts (r-band). We have used the following set of
equations to transform our magnitudes and error(mag) provided by the catalogues into counts and
error(counts):\begin{eqnarray} mag=-2.5\log\left(\frac{counts}{exptime}10^{0.4(aa+kk\times
airmass)}\right), \\ error(mag)=\frac{2.5}{\ln 10}\frac{error(counts)}{counts},\end{eqnarray} with
exptime=53.907456 s and aa (zero-point), kk (extinction coefficient) and airmass provided for each
object.}. Those galaxies in our catalogue which show photometric errors larger than those values (in
any of the two bands) are discarded from the analysis as their large errors could be linked to
artifacts in the image: proximity to bright nearby companions, etc. The number of galaxies rejected due
to large photometric errors are: 4.1 per cent in g-band and 4.5 per cent in r-band.

Finally, there is a number of satellites candidates which are bright enough (r$<$17.7 mag) to have a spectroscopic
redshift determination. Consequently, we can divide our analysis of the satellite galaxies in two subsamples: one
where the redshift of the satellites has been determined spectroscopically and one where we have used the
photometric redshift determination. In the following subsections we explore what are the characteristics of each
subsample.

\subsubsection{Stellar mass completeness for the spectroscopic catalogue}
\label{sub:compspeccat}

Once the stellar masses of the galaxies of the sample are determined we can estimate down to which
stellar mass our catalogue of galaxies is complete. In this subsection we do such analysis for the
spectroscopic catalogue. We conduct the same exercise for the photometric sample in the following
subsection.

\begin{figure}
\includegraphics[width=1.0\columnwidth,clip=true]{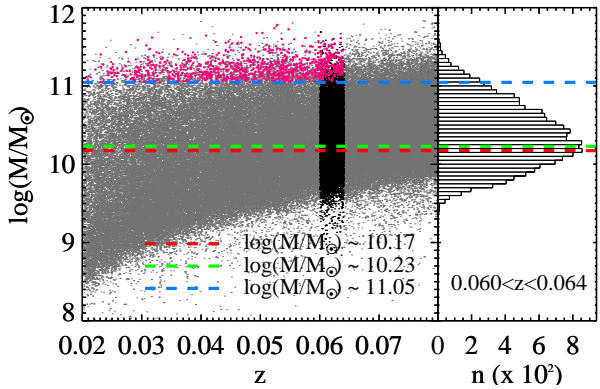}

\caption{Left-hand panel: stellar mass distribution versus redshift for the galaxies in the NYU VAGC
spectroscopic catalogue.  We plot in grey the galaxies found within the spectroscopic catalogue. The
pink points  correspond to our sample of host massive elliptical galaxies. Right-hand panel: stellar mass
distribution  for the spectroscopic catalogue in the redshift interval $0.060<z<0.064$ (plotted in
black in the left-hand panel). This redshift range corresponds to the limiting redshift we have used for the
host galaxies in our spectroscopic sample (see Fig. 1). The  red dashed line is the estimated completeness limit
$\sim$1.5$\times$10$^{10}$M$_\odot$   whereas the blue line represents the minimum stellar mass
estimated for the sample of  massive elliptical galaxies ($\sim$1.1$\times$10$^{11}$M$_\odot$). A 
conservative estimate for the stellar mass completeness of the spectroscopic sample is provided by the
green dashed line: $\sim$1.7$\times$10$^{10}$M$_\odot$ (see text for details).} 

\label{spec_comp} 
\end{figure}

In the left-hand panel of Fig.~\ref{spec_comp}, we present the stellar mass versus the spectroscopic redshift for the
galaxies in the NYU VAGC. In order to estimate the mass completeness  of this sample,  we have explored
the mass  distribution of the galaxies at the upper limit of our host galaxies redshift range (i.e.
0.06$<$z$<$0.064). This is shown in the right-hand panel of  Fig.~\ref{spec_comp}. The mass completeness limit is given
by the position of the peak of the mass distribution. We estimate that peak evaluating the mode of the distribution.
The stellar mass completeness limit for our spectroscopic catalogue is placed on $\log(M_\star/M_\odot$)$\sim$10.17 
at z=0.064. Above this mass value, we can study the satellite population with completeness up to z=0.064. This value
of mass corresponds roughly to a mass ratio of  1:10 ($0.1 < M_{\rm Sat}$ $/M_{\rm Host}< 1$) for the satellite
population.

\subsubsection{Stellar mass completeness for the photometric catalogue}
\label{sub:compcat}

\begin{figure}
\includegraphics[width=1.0\columnwidth,clip=true]{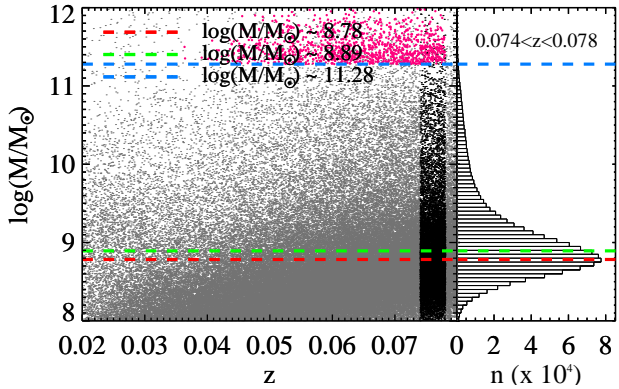}
\caption{Left-hand panel: stellar mass distribution versus redshift for the galaxies in the
photometric photo-z SDSS DR7 catalogue. The galaxies of the photometric sample are represented
in grey. The pink distribution corresponds to our sample of host massive elliptical galaxies.
Right-hand panel: histogram of the stellar mass distribution of galaxies in the photometric
catalogue in the redshift interval $0.074<z<0.078$ (plotted in black in the left-hand panel). This redshift range corresponds to the limiting redshift we have used for the
host galaxies in our photometric sample (see Fig. 1). The  red dashed line is
the estimated completeness limit $\sim$6$\times$10$^{8}$M$_\odot$   whereas the blue
line represents the minimum stellar mass estimated for the sample of  massive elliptical galaxies
($\sim$1.9$\times$10$^{11}$M$_\odot$). A  conservative estimate for the stellar mass completeness of the
spectroscopic sample is provided by the green dashed line: $\sim$8$\times$10$^{8}$M$_\odot$ (see text
for details). We only show a randomly selected 1.5 per cent of the total number of galaxies to avoid overloading
the figure.}
\label{photo_comp} 
\end{figure}

We now conduct the same completeness analysis but  this time using the photo-z SDSS DR7 photometric catalogue. In the
left-hand panel of Fig.~\ref{photo_comp}, we present the distribution of the stellar mass respect to redshift of the
galaxies in the photometric catalogue. As we did with the spectroscopic catalogue, we explore the distribution of
the stellar masses of our galaxies up to the upper limit of our redshift distribution (in this case z=0.078). The
right-hand panel of Fig.~\ref{photo_comp} shows the stellar mass distribution within the redshift interval
0.074$<$z$<$0.078. The stellar mass completeness limit (estimated as the mode of the distribution in that redshift
interval) is established in $\log(M_{\star}/M_\odot$)$\sim$8.78. This value indicates that we can explore with
completeness the distribution of satellites around our host massive ellipticals down to a mass ratio of  $\sim$1:330
(i.e. $0.003<M_{\rm Sat}/M_{\rm Host}<1.0$).

It is worth noting, however, that both in the spectroscopic and the photometric catalogues, there is a
potential bias to miss the oldest galaxies at a fixed stellar mass. This is because both catalogues are
complete in redshift down to a given apparent r-band magnitude. In the case of the spectroscopic
catalogue this is r$\sim$17.7 (90 per cent completeness) and in the case of the photometric sample is
r$\sim$21.5 (95 per cent completeness). These numbers translate into the following K-corrected absolute
magnitude values for each of the samples at z=0.064: M$_{\rm r}$=-19.7 mag in the case of the
spectroscopic sample and  M$_{\rm r}$=-16.4 mag at z=0.078 for the photometric catalogue. To transform
these absolute magnitude values into stellar mass limit we need to have an estimation of the stellar
mass-to-light ratio of our galaxies.  We have estimated which is the average g-r color for our host
galaxies and for our potential satellites. We find that our host ($\sim$10$^{11}$M$_\odot$) galaxies
have a typical (g-r)$\sim$0.85 whereas our less massive ($\sim$10$^{9}$M$_\odot$) galaxies are bluer
(g-r)$\sim$0.70. Let's assume now a very conservative age for the less massive galaxies of 12 Gyr.
Using the MIUSCAT spectral energy distributions (SEDs) developed by \citet{Vazdekis2012} and \citet{Ricciardelli2012}, with the above
color and age and a Kroupa IMF, the largest (M/L)$_{\rm r}$ ratio for these objects will be around 3.
This translates into the following stellar mass figures for our stellar mass completeness:
$\log(M_\star/M_\odot$)$\sim$10.23 for the spectroscopic catalogue (i.e. a mass ratio of 1:7) and
$\log(M_\star/M_\odot$)$\sim$8.89 for the photometric catalogue (i.e. a mass ratio of 1:250). On what
follows, we will consider these values as the most conservative mass completeness limits of our
satellite galaxies. 

\section{Satellite selection criteria}\label{sec:selection}

Both for the spectroscopic and photometric catalogues we have applied the following procedure for identifying
the satellite galaxies around our host objects.


(i) We detect all the galaxies in the SDSS catalogues which are within a projected radial distance to
our central galaxies of R=100 kpc. This corresponds to a radius of 4.1 and 1.13 arcmin at z=0.02 and
z=0.078, respectively. To avoid any bias due to the borders of the SDSS survey, we only have considered
host galaxies such as the area enclosed by the satellite's search radius is fully contained within the
catalogue borders. Our adopted search radius of 100 kpc is a compromise between having a large area for
finding a significant number of satellite candidates gravitationally bound to our central massive
galaxies but not as large as to be severely contaminated by background and foreground objects (see
Section~\ref{sub:background}).

\begin{figure*}
\includegraphics[width=2.0\columnwidth,clip=true]{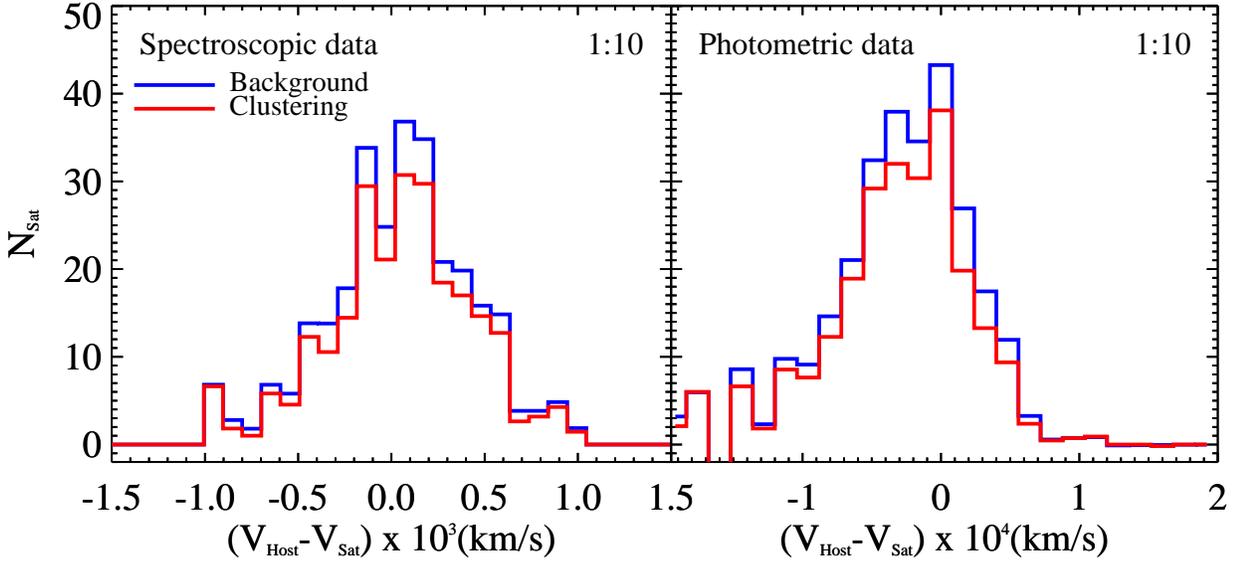}

\caption{Velocity difference distributions between the satellites and the host galaxies within the
spectroscopic and photometric catalogues. This figure illustrates for both cases the distribution in
velocity of satellites with mass down to 1:10 of the host galaxy. The velocity distribution has been
corrected statistically of the expected contamination due to the background and clustering effects. The
width of the spectroscopic distribution reflects the intrinsic velocity dispersion of the satellites
around the host galaxies. The width of the photometric distribution, however, is given ultimately by
the photometric redshift error at measuring the velocity of the galaxies.}

\label{sat_vel} 
\end{figure*}

(ii) The absolute difference between the satellite redshifts and the redshift of the central galaxies
must be lower than 0.0033 (in the case of the spectroscopic catalogue) or 0.066 (for the photometric
catalogue). The spectroscopic $\mid$$\Delta$z$\mid$=0.0033 value was chosen to select only those
objects that are at less than 1000 km/s away from the galaxy host. This value has been used before in
the literature to select gravitationally bound satellites of massive galaxies, see e.g.
\citet{Wang2012}. In order to check whether this criteria is reasonable for our work, in Fig.
\ref{sat_vel} we show the difference in velocity between the hosts and the satellite galaxies selected
with the above constraints. Fig. \ref{sat_vel} illustrates that the velocity distribution is close to a
Gaussian shape with a dispersion of $\sim$300 km/s for the spectroscopic sample. Note that the
individual spectroscopic errors are very small compared to the velocity dispersion of the sample (i.e.
0.0001 in redshift or 30 km/s). Consequently, the vast majority of the satellites of the massive
galaxies are enclosed within our velocity criteria. In the photometric case, the velocity distribution
around the host galaxies is wider as the uncertainty in the velocity of the galaxies is larger.
Following a similar criteria to the one used in the spectroscopic case, we take all the galaxies within
3$\sigma$ of the mean as potential satellite candidates. The sigma of the velocity distribution we have
measured is $\sim$6600 km/s (i.e. $\mid$$\Delta$z$\mid$=0.022). Not surprisingly, this width is
equivalent to the photometric redshift error we have reported previously. Consequently, we take
$\mid$$\Delta$z$\mid$=0.066 as the width of the redshift interval for identifying satellites for the
photometric case. Both the spectroscopy and photometric velocity distributions shown in Fig.
\ref{sat_vel} have been corrected statistically by the effect of background and clustering
contamination. These effects are explained in the next sections.

(iii) The mass ratio
between our host massive galaxy and the potential satellite should be above 1:10 (in the case of the spectroscopic
sample) and 1:400 in the case of the photometric set. 


 Before showing our results, it is worth addressing the potential biases that can
affect our counting of satellites around our massive hosts. This is done in the following subsections.

\subsection{Background estimation}\label{sub:background}

Despite we have used spectroscopic/photometric redshift information to select our potential satellite
galaxies, there is still a fraction of objects that satisfy all the above criteria but are not
gravitationally bound to our massive host galaxies. These objects are counted as satellites because the
uncertainties on their redshift estimates include them within our searching redshift range. These
foreground and background objects (hereafter we will use the term background to refer to both of them)
constitute the main source of uncertainty in this kind of studies. Consequently, it is key to estimate
accurately the background contamination in order to statistically subtract its contribution from the
fraction of galaxies hosting satellites.

To estimate the fraction of background sources that contaminates our satellite
samples we have developed a set of simulations. The procedure consists on placing a
number of mock massive galaxies (equal to the number of our host galaxies)
randomly through the volume of the catalogue, conserving the original
characteristic in the stellar mass and redshift for the sample of massive
elliptical galaxies. Once we have placed our mock galaxies through the catalogue,
we count which fraction of these mock galaxies have 'satellites' around them
taking into account our criteria of mass, redshift and distance explained in the
above section. This procedure was repeated 2000  times to have a robust
estimation of the fraction of mock galaxies with satellites. We define this
average fraction as $S_{\rm Sim}$. These simulations allow us to estimate the
scatter in the fraction of galaxies that have contaminants and use it as an
estimation of the error of our real measurements.

We consider then this fraction to be representative of the background affecting
our real satellite sample. Taking into account that the observed fraction of 
galaxies with satellites, $F_{\rm Obs}$, is the sum of the fraction of galaxies
with real satellites $F_{\rm Sat}$, plus the fraction of galaxies which have not
satellites but are affected by contaminants ($1-F_{\rm Sat}$)$S_{\rm Sim}$, 
thus, we deduce the following expression \citep{Marmol-Queralto2012}:

\begin{equation}\label{formula_frac}
F_{\rm Sat} = \frac{F_{\rm Obs} - S_{\rm Sim}}{1-S_{\rm Sim}}.
\end{equation}

We show in Tables ~\ref{table_results} and ~\ref{table_results.dif} the results of the background
estimations. Table ~\ref{table_results} shows the results for different  'cumulative' mass bins. In
Table 2, each mass ratio bin is 'differential'. It is worth noting how the effect of the background
increases as we move towards smaller mass ratios. For instance, taking a look to Table
~\ref{table_results.dif}, the fraction of massive galaxies expected to have a fake satellite $S_{\rm
Sim}$ using our search criteria is only $\sim$0.5 per cent  when we explore the mass ratio  0.5$<$M$_{\rm
sat}$/M$_{\rm Host}$$<$1.0 (for the photometric catalogue). However, this fraction can rises up to
$\sim$36 per cent when we probe satellites with 0.0025$<$M$_{\rm sat}$/M$_{\rm Host}$$<$0.005 (for the
photometric case). This is as expected as the number of background sources increases as we explore
fainter and fainter objects. We also note that the effect of the background is less relevant when we
use the spectroscopic sample. In fact, the background is $\sim$20 times higher in the photometric
sample than in the spectroscopic one. Again, this is as expected due to the more restrictive search
criteria at decreasing the error at determining the redshift of the satellites. Nonetheless, we draw
the attention of the readers to the remarkable agreement on the corrected fraction of massive galaxies
with satellites $F_{\rm sat:S}$ both using the spectroscopic and photometric catalogues when exploring
common mass ratios. This suggests that our correction of the background works properly in the
photometric sample.

\begin{table*}
\centering
\caption{Cumulative fraction of massive galaxies with at least one satellite found within
the mass ratio satellite-host. $N_{\rm Host,Sat}$ is the number of massive galaxies in our sample with at least a satellite using our
search criteria.  The total number of observed satellites is given by  $N_{\rm Sat,Obs}$. 
The observed fraction of massive galaxies with at least a satellite is provided by $F_{\rm Obs}$. The estimated fraction of host galaxies
with at least a satellite due to the background contamination is
$S_{\rm Sim}$, and due to the clustering effect is $S_{\rm Clu}$. In the last
two columns, we present the corrected fraction of massive galaxies with at least a satellite
when the correction for (i) the background contamination
($F_{\rm sat:S}$) or (ii) the clustering effect ($F_{\rm sat:C}$) is applied.}

\begin{tabular}{lccccccc}
\hline\hline
$M_{\rm Sat}/M_{\rm Host}$ & $N_{\rm Host,Sat}$ &  $N_{\rm Sat,Obs}$ & $F_{\rm Obs}$ & $S_{\rm Sim}$ & $S_{\rm Clu}$  & $F_{\rm Sat:S} $  & $F_{\rm Sat:C}$\\
\hline
                          &       &       &  & Spectroscopic sample    &   &  & \\
\hline
0.500-1.0&$          26$&$          26$&$     0.026\pm     0.004$&$    0.0003\pm   0.00002$&$    0.0050\pm    0.0003$&$    0.0252\pm    0.0042$&$    0.0207\pm    0.0042$\\
0.200-1.0&$         114$&$         122$&$     0.112\pm     0.007$&$    0.0019\pm   0.00003$&$    0.0247\pm    0.0004$&$    0.1104\pm    0.0070$&$    0.0896\pm    0.0072$\\
0.100-1.0&$         243$&$         289$&$     0.239\pm     0.008$&$    0.0034\pm   0.00004$&$    0.0458\pm    0.0007$&$    0.2364\pm    0.0079$&$    0.2024\pm    0.0082$\\
\hline
                         &       &       &  & Photometric sample    &   &  & \\
\hline
0.5000-1.0&$          45$&$          47$&$    0.0392\pm    0.0047$&$    0.0070\pm    0.0001$&$    0.0140\pm    0.0002$&$    0.0325\pm    0.0047$&$    0.0256\pm    0.0048$\\
0.2000-1.0&$         162$&$         186$&$    0.1412\pm    0.0069$&$    0.0321\pm    0.0003$&$    0.0532\pm    0.0004$&$    0.1128\pm    0.0072$&$    0.0930\pm    0.0073$\\
0.1000-1.0&$         282$&$         365$&$    0.2459\pm    0.0074$&$    0.0655\pm    0.0004$&$    0.1014\pm    0.0007$&$    0.1930\pm    0.0079$&$    0.1607\pm    0.0082$\\
0.0500-1.0&$         421$&$         621$&$    0.3670\pm    0.0071$&$    0.1123\pm    0.0006$&$    0.1707\pm    0.0005$&$    0.2870\pm    0.0079$&$    0.2368\pm    0.0085$\\
0.0200-1.0&$         621$&$        1115$&$    0.5414\pm    0.0057$&$    0.2080\pm    0.0008$&$    0.2909\pm    0.0010$&$    0.4210\pm    0.0072$&$    0.3532\pm    0.0081$\\
0.0100-1.0&$         782$&$        1648$&$    0.6818\pm    0.0042$&$    0.3349\pm    0.0009$&$    0.4266\pm    0.0014$&$    0.5215\pm    0.0064$&$    0.4450\pm    0.0074$\\
0.0050-1.0&$         929$&$        2382$&$    0.8099\pm    0.0027$&$    0.5035\pm    0.0009$&$    0.5993\pm    0.0011$&$    0.6172\pm    0.0054$&$    0.5257\pm    0.0066$\\
0.0025-1.0&$        1027$&$        3162$&$    0.8954\pm    0.0015$&$    0.6522\pm    0.0009$&$    0.7403\pm    0.0009$&$    0.6992\pm    0.0043$&$    0.5972\pm    0.0058$\\

\hline\hline
\end{tabular}
\label{table_results}
\end{table*}

\begin{table*}
\centering
\caption{Differential fraction of massive galaxies with at least one satellite found within
the mass ratio satellite-host. $N_{\rm Host,Sat}$ is the number of massive galaxies in our sample with at least a satellite using our
search criteria.  The total number of observed satellites  is given by  $N_{\rm Sat,Obs}$. 
The observed fraction of massive galaxies with at least a satellite is provided by $F_{\rm Obs}$. The estimated fraction of host galaxies
with at least a satellite due to the background contamination is
$S_{\rm Sim}$, and due to the clustering effect is $S_{\rm Clu}$. In the last
two columns, we present the corrected fraction of massive galaxies with at least a satellite
when the correction for (i) the background contamination
($F_{\rm sat:S}$) or (ii) the clustering effect ($F_{\rm sat:C}$)  is applied .}

\begin{tabular}{lccccccc}
\hline\hline
$M_{\rm Sat}/M_{\rm Host}$ & $N_{\rm Host,Sat}$ &  $N_{\rm Sat,Obs}$ & $F_{\rm Obs}$ & $S_{\rm Sim}$ & $S_{\rm
Clu}$  & $F_{\rm Sat:S} $  & $F_{\rm Sat:C}$\\
\hline
              &   &  &      &      Spectroscopic sample    &   &  & \\
\hline
0.500-1.000&$          26$&$          26$&$     0.026\pm     0.004$&$    0.0003\pm   0.00002$&$    0.0050\pm    0.0003$&$    0.0252\pm    0.0042$&$    0.0207\pm    0.0042$\\
0.200-0.500&$          93$&$          96$&$     0.091\pm     0.007$&$    0.0014\pm   0.00003$&$    0.0202\pm    0.0004$&$    0.0902\pm    0.0066$&$    0.0727\pm    0.0068$\\
0.100-0.200&$         148$&$         167$&$     0.146\pm     0.007$&$    0.0015\pm   0.00003$&$    0.0223\pm    0.0004$&$    0.1443\pm    0.0074$&$    0.1261\pm    0.0076$\\
\hline
                   &   &  &  &     Photometric sample    &   &  & \\
\hline
0.5000-1.000&$          45$&$          47$&$    0.0392\pm    0.0047$&$    0.0070\pm    0.0001$&$    0.0140\pm    0.0002$&$    0.0325\pm    0.0047$&$    0.0256\pm    0.0048$\\
0.2000-0.500&$         126$&$         139$&$    0.1099\pm    0.0065$&$    0.0254\pm    0.0002$&$    0.0409\pm    0.0004$&$    0.0867\pm    0.0067$&$    0.0719\pm    0.0068$\\
0.1000-0.200&$         156$&$         179$&$    0.1360\pm    0.0069$&$    0.0363\pm    0.0003$&$    0.0532\pm    0.0005$&$    0.1035\pm    0.0071$&$    0.0875\pm    0.0073$\\
0.0500-0.100&$         222$&$         256$&$    0.1935\pm    0.0073$&$    0.0539\pm    0.0003$&$    0.0806\pm    0.0008$&$    0.1476\pm    0.0077$&$    0.1228\pm    0.0079$\\
0.0200-0.050&$         374$&$         494$&$    0.3261\pm    0.0072$&$    0.1167\pm    0.0005$&$    0.1523\pm    0.0009$&$    0.2370\pm    0.0082$&$    0.2049\pm    0.0085$\\
0.0100-0.020&$         416$&$         533$&$    0.3627\pm    0.0071$&$    0.1740\pm    0.0006$&$    0.2040\pm    0.0009$&$    0.2284\pm    0.0086$&$    0.1994\pm    0.0089$\\
0.0050-0.010&$         534$&$         734$&$    0.4656\pm    0.0064$&$    0.2783\pm    0.0007$&$    0.3186\pm    0.0007$&$    0.2595\pm    0.0089$&$    0.2157\pm    0.0094$\\
0.0025-0.005&$         554$&$         780$&$    0.4830\pm    0.0063$&$    0.3312\pm    0.0007$&$    0.3644\pm    0.0015$&$    0.2270\pm    0.0094$&$    0.1866\pm    0.0098$\\

\hline\hline
\end{tabular}
\label{table_results.dif}
\end{table*}

\subsection{Clustering effect}\label{sub:clustering}

As we are dealing with nearby massive elliptical galaxies which tend to populate regions with
an evident overdensity compared to the average density of the Universe, it is worth
exploring whether our background correction is representative of the contamination of sources
surrounding our host galaxies. The overdense environment  leads to an excess of probability
(which we term as clustering) of finding galaxies that could be misidentified as satellites
of our selected sources.  Even with all the redshifts measured spectroscopically, the effect
of clustering is relevant in our estimates since this effect is inherent to our inability of
measuring real distances in regions such as galaxy clusters where the velocities of the
galaxies depart from a pure Hubble flow significantly. Therefore, in cluster of galaxies the
velocity dispersion of the cluster will limit ultimately our accuracy on estimating real
galaxy associations. 

\subsubsection{How do we estimate the clustering?}\label{sub:clustestimates}

The clustering is a local effect affected by the substructure of the clusters. Consequently,
one would like to measure its influence as close as possible to the host galaxies. In
practice, this is done by measuring the number of satellite candidates in different annuli
beyond our search radius \citep{Chen2006, Liu2011, Marmol-Queralto2012}. We will call S$_{\rm
Clu}$ to the fraction of massive galaxies having 'satellites'  satisfying our selection
criteria within these external annuli.  This fraction measures both the effect of the
background contamination plus the excess over this background due to the clustering. This
method has the disadvantage, compared to the simulations that we have conducted above, that
is statistically more uncertain since S$_{\rm Clu}$ can be measured only around our massive
galaxies and this number is relatively small.

\begin{figure}
\centering
\includegraphics[width=1.0\columnwidth,clip=true]{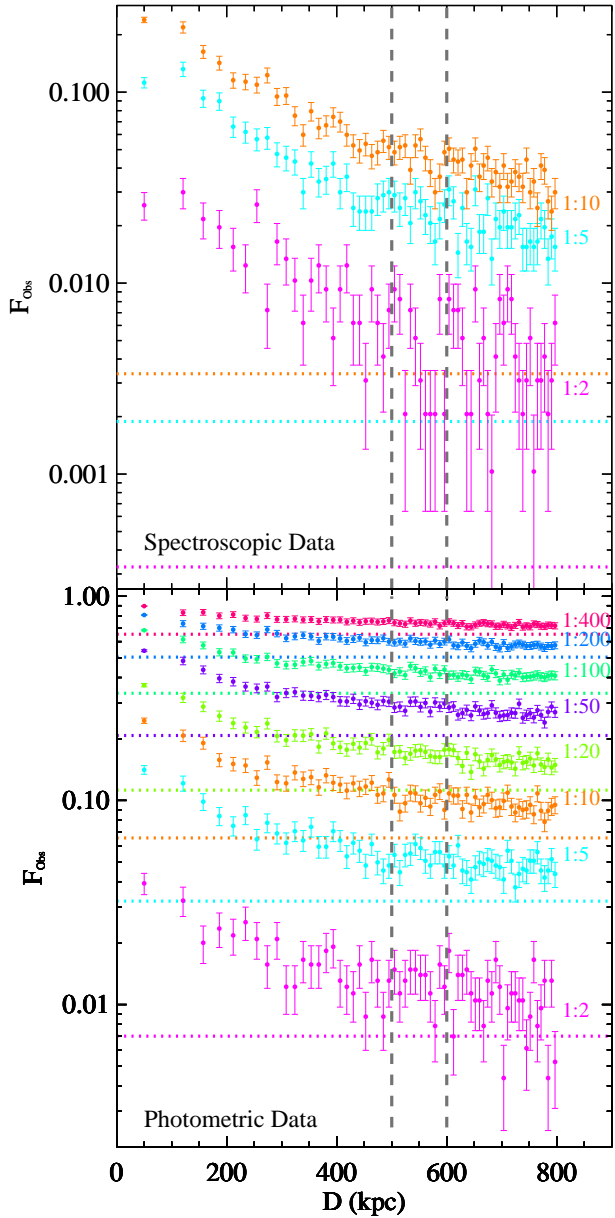}

\caption{Upper panel: observed fraction of massive elliptical galaxies with at least a satellite versus
 the projected radial distance satellite-host for 64 equal area rings. The colours represent the
 different mass ratio (M$_{\rm Sat}$/M$_{\rm Host}$) bins explored in this work.  The coloured dashed
 lines correspond to the expected fraction of massive elliptical galaxies with at least a fake
 satellite due to background contamination (i.e. $S_{\rm Sim}$; see Section~\ref{sub:background}). The
 vertical lines enclose the region where the clustering has been measured. Lower panel: same than above
 but using the photometric catalogue instead of the spectroscopic data.  Errors bars for each ring are
 derived assuming Poissonian noise.}

\label{clust}
\end{figure}

To quantify the clustering, we count the satellites in 64 different annuli in the radial
range 100$<$R$_{\rm Search}$$<$800 kpc, where the size of each annuli was selected to contain
the same area than the main searching aperture (i.e. $\pi$(100 kpc)$^2$). As Fig.
~\ref{clust} illustrates, the observed fraction of massive galaxies with at least a satellite
smoothly declines towards the outer radii. As the radial distance increases, we expect that 
F$_{\rm Obs}$ asymptotically reaches the background value (i.e. S$_{\rm Sim}$). However, it
is worth noting that even at distances as far as 800 kpc we do not yet reach the values
obtained in the above background estimation method. This indicates that the effect of the
clustering is significant even at distances as large as 800 kpc, and therefore, we cannot
dismiss the clustering effect in our analysis\footnote{Note that 800 kpc is still well inside
the typical virial radius for massive galaxy clusters which is around 1-2 Mpc. In fact, we
have conducted a simple extrapolation of the data shown in Fig.\ref{clust} and found that
F$_{\rm Obs}$ asymptotically reaches the background value at D$\sim$2 Mpc.}.

The next step is then to decide at which radial distance from the host galaxies should we estimate
$S_{\rm Clu}$. We have finally chosen to define S$_{\rm Clu}$ as the average value of  F$_{\rm Obs}$ in
the 11 annuli between 500 and 600 kpc (see  Fig.~\ref{clust}). This radial range is a compromise among
having a local measurement of the environment around our massive host galaxies but being  far away
enough such as the probability of finding a gravitationally bounded satellite to our targeted galaxy
will be low. The  projected radial distance of 500 kpc is chosen following many works in the literature
\citep[e.g.][]{Sales2004,Sales2005,Chen2006,Bailin2008,Wang2012} which have used only galaxies with
radial distances less than 500 kpc to define their sample of truly (i.e. bounded) satellite galaxies.
The clustering values that we have estimated are compiled in Tables ~\ref{table_results} and
~\ref{table_results.dif}. To estimate the uncertainty at determining $S_{\rm Clu}$ for each mass ratio,
we took advantage of the gentle decline of  $F_{\rm Obs}$ with the radial distance. We have fitted
with a straight line the 11 data points within 500 and 600 kpc and then we have measured the rms
respect to the fit. That rms is the quoted uncertainty.

Finally, it is worth comparing the obtained values for $S_{\rm Clu}$ at using both the
spectroscopic and the photometric catalogues in the common mass ratio bins. Contrary to the
background contamination, the clustering effect is dominated by the velocity dispersion of
the clusters. If as we expect, the clustering effect is  mainly related with our inability of
measuring real distances due to the velocity dispersion of  the clusters, S$_{\rm Clu}$
should be similar independently of the redshift catalogue used. The values provided in Tables
~\ref{table_results} and ~\ref{table_results.dif} indicate that this is in fact the case. Our
results show that the spectroscopic clustering value is only slightly smaller (a factor of
$\sim$2) than  the photometric one. Another important aspect to note is that being S$_{\rm
Clu}$ systematically larger than S$_{\rm Sim}$, the corrections in the observed fraction of
massive galaxies with a least a satellite are larger.

\section{Results}\label{results}

\subsection{Fraction of massive ellipticals having satellites}

\begin{figure}
\includegraphics[width=1.00\columnwidth,angle=0,clip=true]{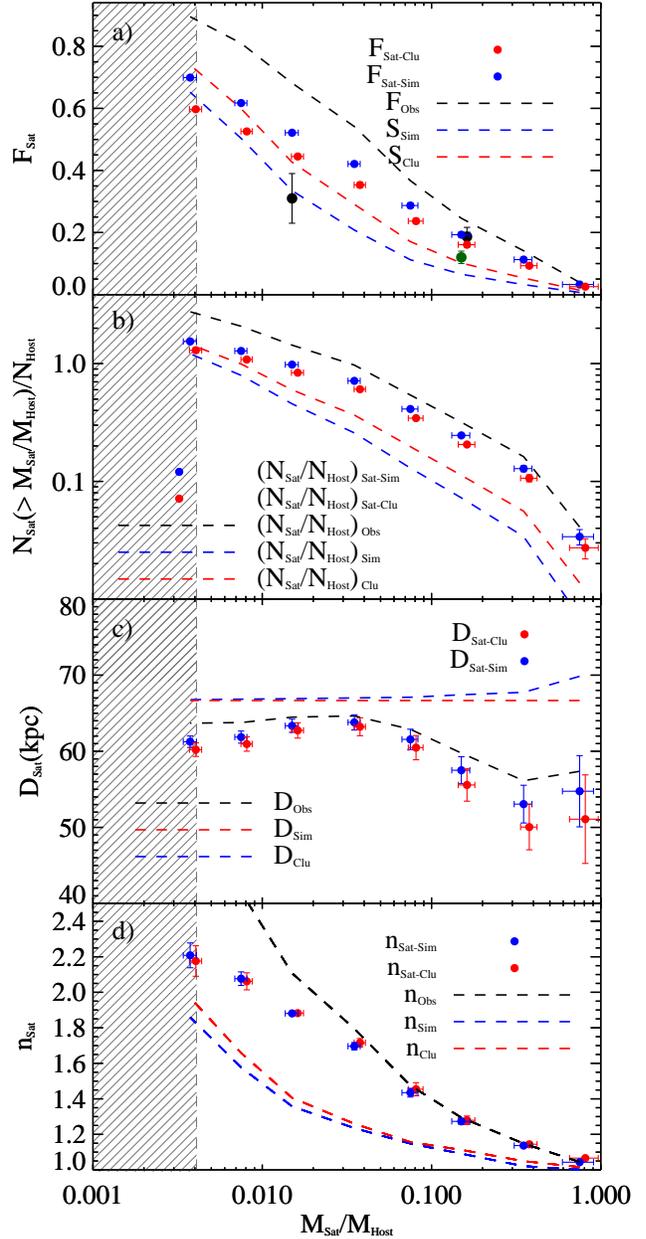}

\caption{Panel a): cumulative fraction of massive elliptical galaxies which host at least one
satellite as a function of the mass ratio M$_{\rm Sat}$/M$_{\rm Host}$ for the photometric
sample.  Points plotted in red show the fraction of massive elliptical galaxies with
satellites after applying the clustering correction whereas the blue points are the result of
correcting by the background. The dashed lines correspond to  $F_{\rm Obs}$ (black), $S_{\rm
Sim}$ (blue) and $S_{\rm Clu}$ (red) (see Sections~\ref{sub:background} and
~\ref{sub:clustering}).  The black solid points correspond to the results by
\citet{Marmol-Queralto2012} for 1:10 and 1:100 mass ratios. The fraction of Milky Way (MW)-like
galaxies with satellites obtained by \citet{Liu2011} is represented by a green point for
satellites down to a 1:10 mass ratio. The vertical grey dashed area corresponds to the mass
ratio region where incompleteness in our photometric satellite sample could play a role (see
a discussion in Section~\ref{sub:compcat}). Panel b): average cumulative number of satellites per galaxy
host  vs. the mass ratio satellite-host. Panel c): projected average radial distance of the
satellites to their galaxy hosts.  Panel d): cumulative multiplicity of satellites around the
massive ellipticals as a function of the mass ratio satellite-host.} 

\label{alldata} 
\end{figure}

Table ~\ref{table_results} and Fig. \ref{alldata} show the cumulative fraction of massive
elliptical galaxies that host satellites within a projected radial distance of 100 kpc as a
function of the mass ratio M$_{\rm Sat}$/M$_{\rm Host}$. In Fig. \ref{alldata}, for our
photometric sample, we show the
observed value, the expected contaminated fraction due to the background, the clustering
effects and  the final corrected fraction of massive galaxies hosting at least one satellite
when the observed data is corrected of these effects.

According to Fig. \ref{alldata}, the cumulative fraction of massive ellipticals hosting at
least one satellite grows almost linearly with the logarithm of the mass ratio between the
satellite and the host. This result holds independently of the correction applied (either 
background or clustering) although with a different slope as expected. The grey vertical area
corresponds to the mass bin where our photometric satellite sample could be incomplete.

\begin{figure}
\includegraphics[width=1\columnwidth,clip=true]{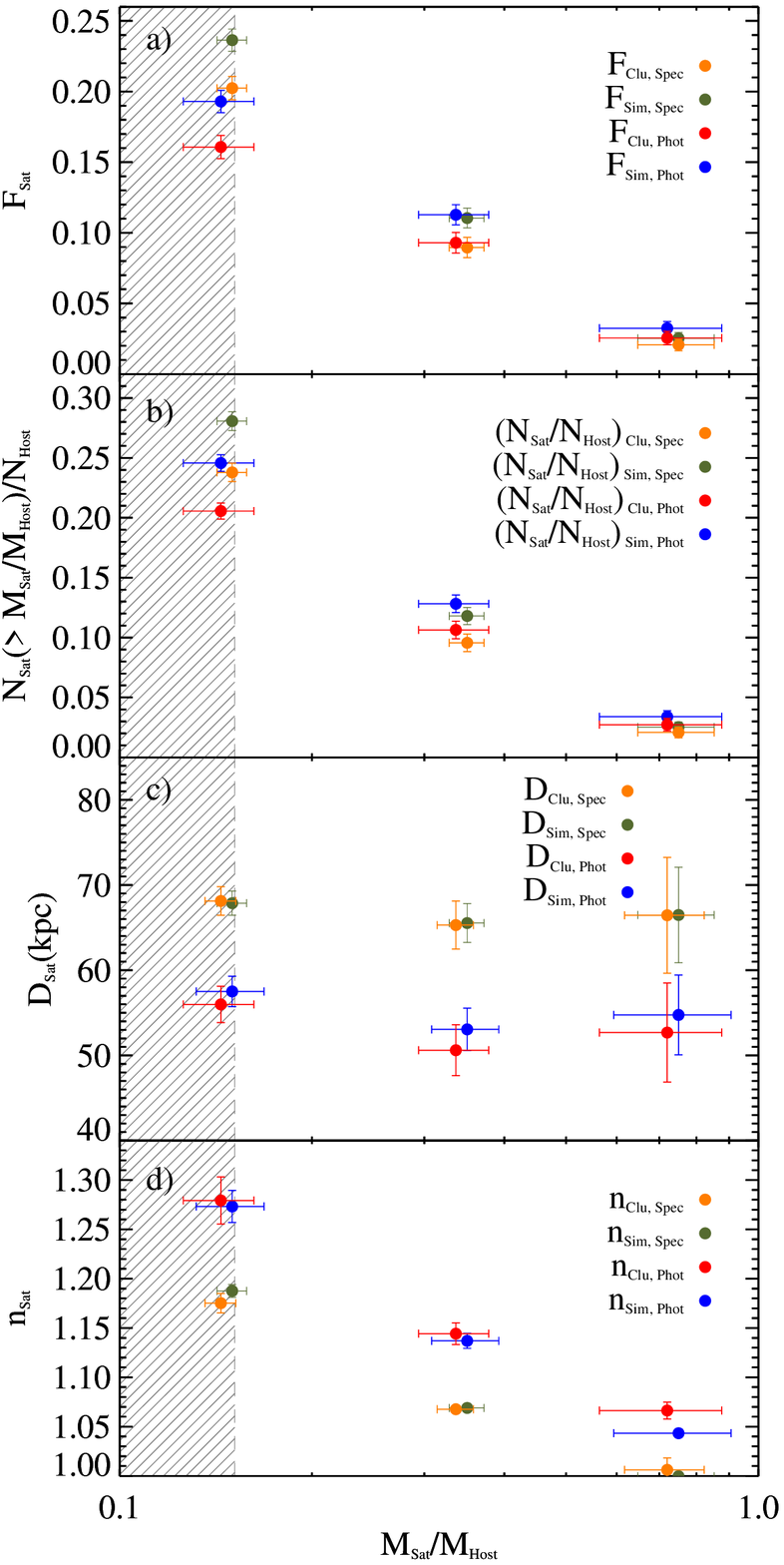}

\caption{Panel a): comparison of the cumulative fractions of massive galaxies with satellites
versus the mass ratio satellite-host (1:2, 1:5, 1:10) for  the photometric and spectroscopic
catalogues once applied the corrections of background and clustering. Panel b): average
cumulative number of satellites per galaxy host versus the mass ratio satellite-host. Panel c):
projected average radial distance of the satellites to their galaxy hosts.  Panel d):
cumulative multiplicity of satellites around the massive ellipticals as a function of the
mass ratio satellite-host. The vertical grey dashed area corresponds to the mass ratio region
where incompleteness in the spectroscopic satellite sample could play a role (see a
discussion in Section~\ref{sub:compcat}). The colour code of each data point is specified in the legend.}

\label{redshiftgraphs_spec}
\end{figure}

It is worth also checking whether the cumulative fraction of massive ellipticals hosting at
least one satellite is comparable when we use either the photometric or the spectroscopic
catalogues. This is done in Fig. ~\ref{redshiftgraphs_spec}. Note the nice agreement between
the values got using the two different catalogues. This shows the consistency of our results
using different samples. 

\begin{figure}
\includegraphics[width=1.00\columnwidth,angle=0,clip=true]{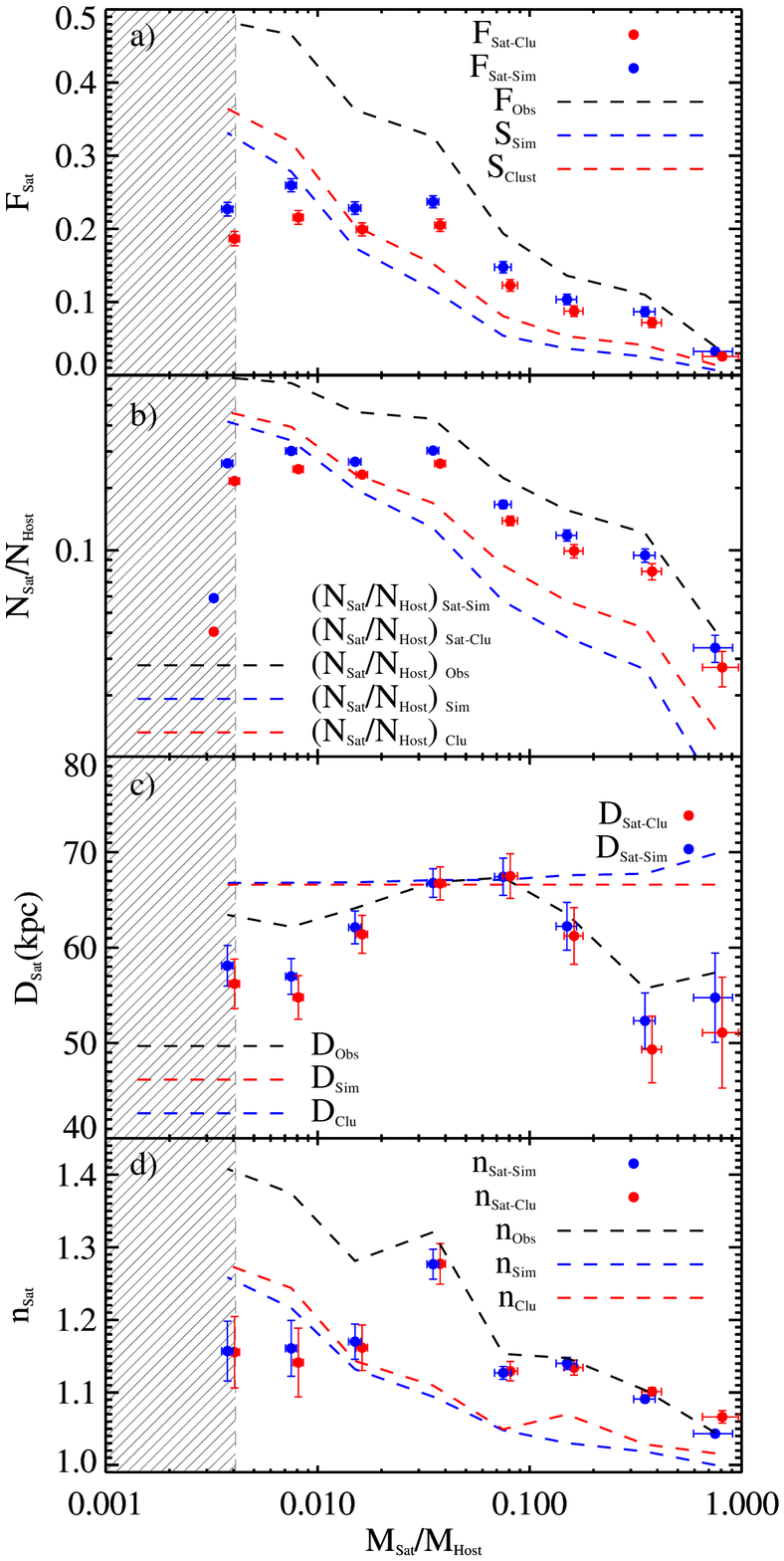}

\caption{Panel a): fraction of massive elliptical galaxies with satellites  versus the mass
ratio  M$_{\rm Sat}$/M$_{\rm Host}$ for the photometric sample. Data plotted in red show the
fraction of massive elliptical galaxies with satellites after applying the clustering
correction whereas the blue points correspond to correction using the background. The dashed
lines show  $F_{\rm Obs}$ (black),  $S_{\rm Sim}$ (blue) and $S_{\rm Clu}$ (red) (see
Sections~\ref{sub:background} and ~\ref{sub:clustering}). The vertical grey dashed area
corresponds to the mass ratio region where incompleteness in the photometric satellite sample
could play a role (see a discussion in Section~\ref{sub:compcat}). Panel b): average number of satellites
per galaxy host  versus the mass ratio satellite-host. Panel c): projected average radial
distance of the satellites to their galaxy hosts.  Panel d):  multiplicity of satellites
around the massive ellipticals as a function of the mass ratio satellite-host.} 

\label{alldata.dif} 
\end{figure}

We can now repeat the same exercise as above but showing which fraction of massive ellipticals has at
least one satellite at different mass ratio bins. The bins we have chosen are shown in Table
~\ref{table_results.dif} and Fig. ~\ref{alldata.dif}. One of the most remarkable results we found in
this paper is that contrary to what we see in the case of the cumulative mass ratio, when we explore
the differential mass ratio, the fraction of massive galaxies with at least one low mass (i.e. with
M$_{\rm Sat}$/M$_{\rm Host}$$<$0.05) satellite remains almost constant ($\sim$20 per cent). We will discuss
the implication of this result in the discussion section.

\subsection{Properties of the population of satellite galaxies}

In addition to quantify the fraction of massive elliptical galaxies having satellites, for each mass
ratio M$_{\rm Sat}$/M$_{\rm Host}$, we estimate the average projected radial distances of these
satellites to their host $D_{\rm Sat}$, the average number of satellites per galaxy host (i.e. N$_{\rm
Sat}$/$N_{\rm Host}$), and the typical number of companions found around the central objects in those
cases where they have at least one satellite $n_{\rm Sat}$ (i.e. the multiplicity). To estimate
properly these quantities, we need to correct for the effect of the contaminants. To achieve this goal,
we have carried on the following procedure. We start with the equality:

\begin{equation}\label{formula_properties_1}
N_{\rm Obs} = N_{\rm Sat} + N_{\rm cont},
\end{equation}

where $N_{\rm Obs}$ is the observed number of satellite candidates, $N_{\rm Sat}$  is the
number of real satellites and $N_{\rm cont}$  is the number of galaxies  missidentified as
satellites (i.e. contaminants). These fake satellites should be represented by the typical
number of contaminants found in the simulations  ($N_{\rm cont}\sim N_{\rm Sim}$) already
introduced in the section~\ref{sub:background}\footnote{To illustrate the methodology, we
refer only here to the background correction, but exactly the same procedure applies for
correcting the clustering effect by simply changing $N_{\rm Sim}$ by $N_{\rm Clu}$.}.
Rewriting Eq.~\ref{formula_properties_1} using probability distributions  at a given radial
distance R  (i.e. $N_{\rm k}(R)= P_{\rm k}(R) N_{\rm Total,k}$) we obtain

\begin{equation}\label{formula_properties_2}
P_{\rm Sat}(R) N_{\rm Sat} = P_{\rm Obs} (R) N_{\rm Obs} - P_{\rm Sim}(R) N_{\rm Sim},
\end{equation}

We can now exemplify this technique. Let us define the average projected radial distance of the
satellites as

\begin{equation}\label{formula_properties_3}
<D_{\rm k}>= \int_0^R P_{\rm k}(R')R'\,dR',
\end{equation}

we get the following equation:

\begin{equation}\label{formula_properties_4}
<D_{\rm Sat}>= \frac{N_{\rm Obs}}{N_{\rm Obs}-N_{\rm Sim}} <D_{\rm Obs}> - 
\frac{N_{\rm Sim}}{N_{\rm Obs}-N_{\rm Sim}} <D_{\rm Sim}> ,
\end{equation}

$<D_{\rm Sat}>$ is the average projected radial distance of the real satellite galaxies after the background
correction. $<D_{\rm Obs}>$ is the average projected radial distance measured directly in the data before the
background correction and  $<D_{\rm Sim}>$ is the average projected radial distance found in  the background
simulations. Other properties such as the average mass fraction of the satellites, their colors, etc. could be
evaluated using the above expression simply replacing the  average radial distances by the average quantity that we
would like to estimate.

\subsubsection{Average projected radial distance satellite-host}\label{sub:distance}

The average projected radial distance of the satellites to the galaxy host is a measurement about the
vicinity of the satellite population to their main galaxies. A change of this quantity with cosmic
time could be an indication of a progressive infall of the satellites towards their host galaxies. In
this work, we quantify this measurement at z=0, helping future works at higher redshifts to explore
changes of this quantity with cosmic time.

\begin{figure}
\centering
\includegraphics[width=1\columnwidth,clip=true]{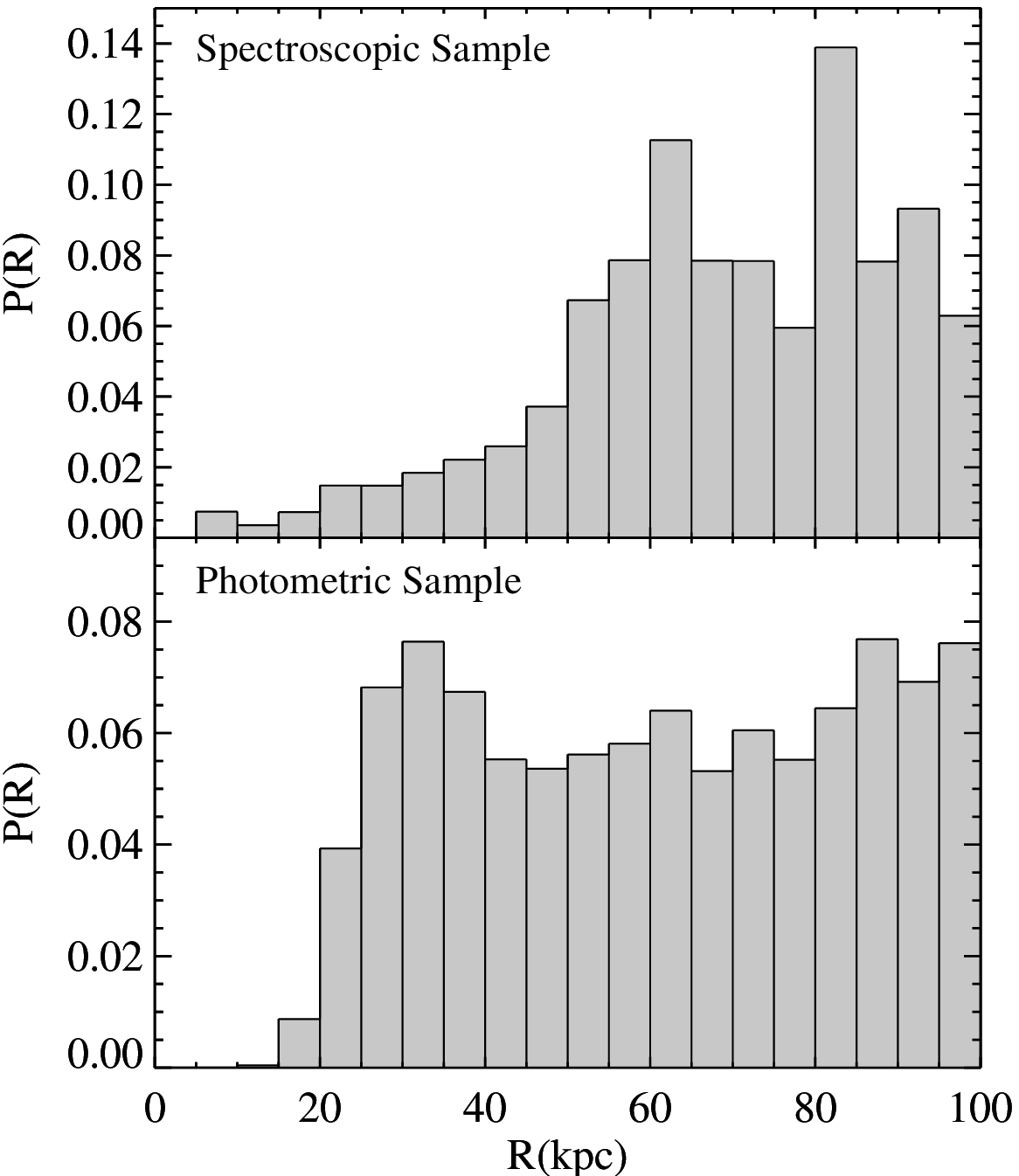}

\caption{Probability distribution of finding a satellite galaxy at a given radial distance to
their hosts once applied the correction of the background contaminants. The upper panel shows
the distribution of the satellite galaxies in the spectroscopic sample down to a mass ratio
of 1:10. The bottom panel shows the distribution of the satellite galaxies in the photometric
sample down to a mass ratio of 1:400. For the photometric sample the hole in the inner region
is due to the presence of the host galaxy. In the spectroscopic case, this hole is caused by
the presence of the host galaxy plus the impossibility of  measuring the spectroscopic
redshifts of two nearby sources at a radial distance closer than 55 arcsec due to  SDSS fiber
collisions.}

\label{splitinmass}
\end{figure}

 The corrected average radial projected distance of
the satellites $<$D$_{\rm Sat}$$>$ are shown in Fig. \ref{alldata} and \ref{alldata.dif} for the
cumulative and differential case, respectively. It is worth noting that in the two cases the average
distance of the galaxies is relatively constant 58.5 kpc (cumulative case) and 59  kpc (differential
case), independently of the explored mass ratio. This value is lower than the expected theoretical
value for a random distribution of objects within a circle of R=100 kpc which is 66.6 kpc. As it is
shown in those figures, the background simulation recovers the expected theoretical value. For the
clustering correction, we have assumed the theoretical value $<$D$_{\rm Clu}$$>$=66.6 kpc since our
method to assess the clustering does not allow us to estimate such quantity directly as we analyze the
rings beyond 100 kpc. The comparison between the photometric and the spectroscopic samples is shown in
panel (c) of Fig. \ref{redshiftgraphs_spec}. Interestingly, the projected distances measured using the
spectroscopic redshift are systematically above the values using the photometric sample. This  result
is likely connected with the fact that two spectroscopic fibers cannot be placed closer than 55 arcsec
on a given plate in the SDSS. At the typical redshift of the spectroscopic sample
(z$\sim$0.049), this distance corresponds to the following rest-frame distance 52.7 kpc. We illustrate
this observational 'hole' of the spectroscopic sample in the upper panel of Fig. \ref{splitinmass}.

The values that we have estimated, consequently, should be seen as upper limits of the real average
radial projected distance of the satellites. This is because we have assumed at correcting our values
that all distances, from 0 to 100 kpc can be occupied by the satellites. In observations, however, the
distribution of satellites does not include objects within the inner $\sim$15-20 kpc in the photometric
sample and very little up to 50 kpc in the spectroscopic sample  (see Fig. \ref{splitinmass}). In the
case of the photometric sample, satellites with smaller projected distances to the host galaxies have
either been cannibalized or their light could be eclipsed by the hosts. In this sense, we will be
unable to separate those objects from the host. In the case of the spectroscopic sample, we have also
added the problem of the fiber collisions as we stated above. 

 How much this effect could affect our distance estimation? We have made a crude estimation of this
effect as follows. Assuming that the probability distribution of the radial distribution of the
satellites P(R) is constant (see e.g. Fig. \ref{splitinmass}), we obtain

\begin{equation}
<D_{\rm Sat}>_{corr}=\frac{1}{1+\alpha}<D_{\rm Sat}>,
\end{equation}

where $\alpha$ is the radial fraction occupied by the host galaxy compared to the total search radius.
In the photometric case $\alpha$=0.15-0.2, which implies that a much reliable value for  $<$D$_{\rm
Sat}$$>$ is 49-51 kpc. In the spectroscopic case we have tentatively assume that $\alpha$ is 0.5,
obtaining $<$D$_{\rm Sat}$$>$$\sim$45 kpc. We will discuss this result as well as the independence with
respect to the mass ratio of the average projected radial distance of the satellites in the discussion
section.

\subsubsection{Average number of satellites per galaxy host}\label{sub:satcandfound}

We also quantify which is the average number of satellites per galaxy host $N_{\rm Sat}/N_{\rm Host}$
as a function of the mass ratio satellite-host. We show this in Figs. ~\ref{alldata} and
~\ref{alldata.dif}. These values have been corrected of the effect of background and clustering. We
have done this by subtracting to $N_{\rm Obs}$ the typical number of contaminants  $N_{\rm Sim}$ and
$N_{\rm Clu}$ found in the background simulation and in our estimates of clustering respectively.

According to Fig.~\ref{alldata} there is only $\sim$0.25 satellites per massive elliptical if we explore
satellites down to a mass ratio 1:10. However, this number grows up to $\sim$1 satellite per massive host if
we explore satellites down to a mass ratio 1:100. That means that, on average, almost all massive ellipticals
(i.e. with a M$_\star$$\gtrsim$10$^{11}$M$_{\sun}$) have a companion with a mass larger than
M$_\star$$\gtrsim$10$^{9}$M$_{\sun}$. This number, however, does not seem to grow fast (i.e. exponentially) when we
explore even lower mass satellites. This is well seen in Fig. ~\ref{alldata.dif}. We see that for satellites
less massive than M$_\star$$\lesssim$10$^{10}$M$_{\sun}$, the average number of satellites per galaxy host in each
mass ratio bin is almost constant. That means that the cumulative number of satellites as we decrease in stellar mass only
grows approximately linearly with  $\log$M$_\star$ of the satellites.

\subsubsection{Multiplicity around the massive elliptical galaxies}\label{sub:numbersat}

To end this section about the properties of the satellite distribution around massive ellipticals, we
explore the multiplicity $n_{\rm Sat}$ of satellites around our massive hosts. In other words, we probe
which is the typical number of satellites in those cases where there is at least one satellite found.

Our findings are shown again in Figs. ~\ref{alldata} and ~\ref{alldata.dif} for the cumulative and the
differential case. For the differential case, our results indicate that at every mass bin (i.e. Fig.
~\ref{alldata.dif}), there is only a small probability of finding more than one satellite within the
same mass ratio. That means that once we have fixed the stellar mass of the satellite, the probability
of finding another one surrounding the massive elliptical with a similar mass is very low. This is
independent of the mass bin explored and does not increase towards satellites less and less massive.
Consequently, when we explore the cumulative multiplicity $n_{\rm Sat}$ (i.e. Fig. ~\ref{alldata}),
we only see a moderate increase towards larger and larger mass ratios. For instance, to get a
cumulative multiplicity larger than 2, we need to considerably decrease the stellar mass of the
satellites probed (i.e. we need to explore satellites at least down to a mass of 
M$_\star$$\sim$10$^{9}$M$_{\sun}$).

\section{The merging channel of massive ellipticals}\label{sub:merging channel}

We now explore a more speculative aspect of our work. Eventually, some satellites surrounding
our massive ellipticals will infall into their massive hosts. Consequently, we can estimate
which are the properties of the satellites that could contribute most to a potential mass
growth of the host galaxies in the future. On what follows, we assume that all satellites,
independently of their mass, will infall with the same speed towards the central galaxy.
Note, however, that this could be not necessary true, as it is theoretically expected that
larger the mass of the satellite shorter will be its merging timescale \citep{Jiang2013}.

To probe the most likely merging channel what we have done is the following. We have estimated all the stellar mass
that is contained by the satellites of a given mass ratio in our sample and we have divided this quantity by the sum
of the mass of all the host galaxies. In other words, we have calculated the following quantity:

\begin{equation}
\Psi=\frac{\sum\limits_{i=1}^{N_{\rm Sat-bin}}M_{\rm Sat-bin,i}}{\sum\limits_{j=1}^{N_{\rm Host}} M_{\rm Host,j}}.
\end{equation}

The sum of all the mass in the host galaxies is a fixed quantity for our samples and their values are 
$\sum_{j=1}^{N_{\rm Host}} M_{\rm Host,j}$=1.8$\times$10$^{14}$M$_{\sun}$ (spectroscopic sample) and $\sum_{j=1}^{N_{\rm
Host}} M_{\rm Host,j}$=4.1$\times$10$^{14}$M$_{\sun}$ (photometric sample). 

\begin{figure}

\includegraphics[width=1\columnwidth,clip=true]{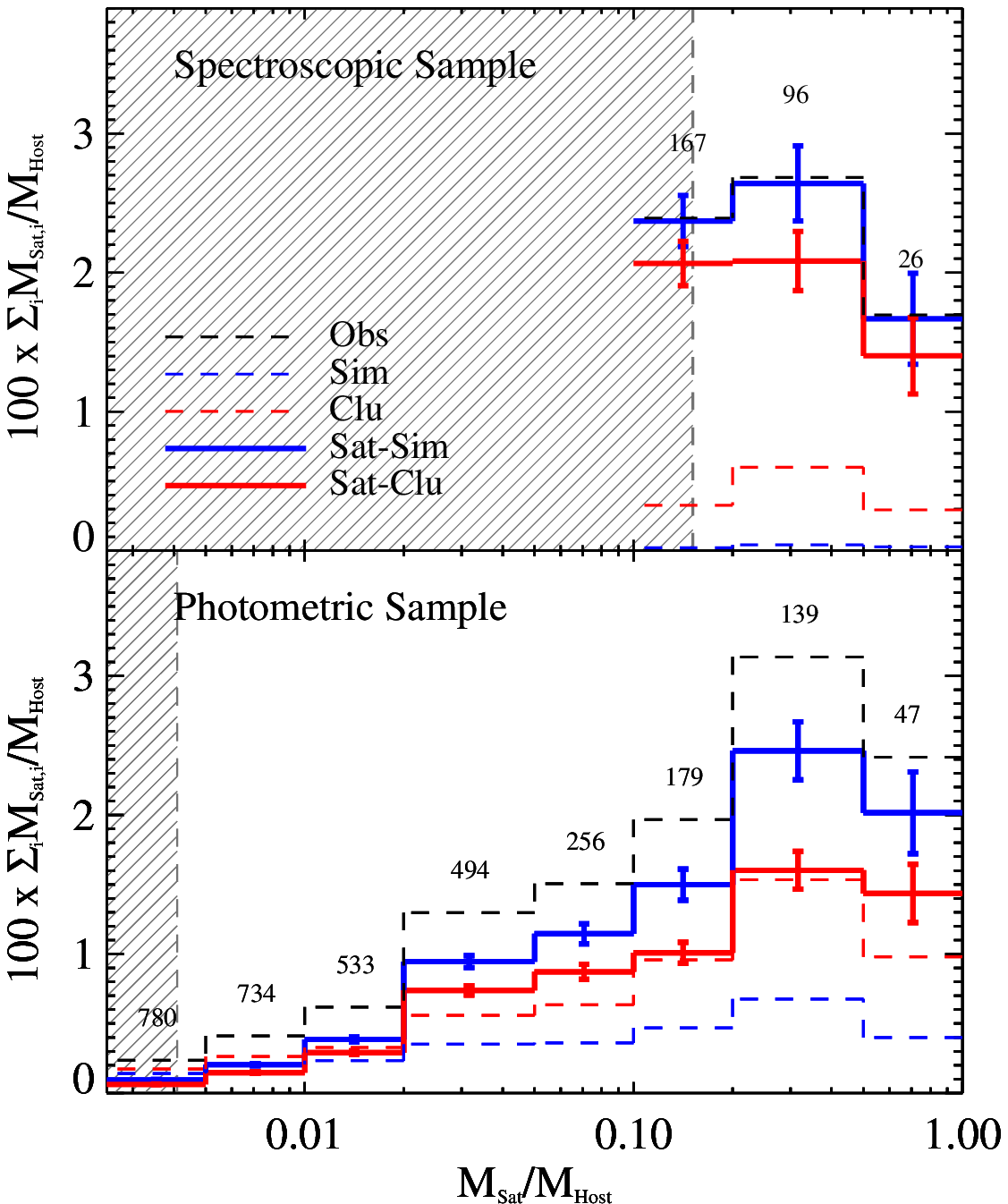}

\caption{The merging channel of massive elliptical galaxies. The figure shows the
contribution of the satellite mass enclosed in each satellite mass bin to the total mass
confined in their hosts. The red solid line represents this quantity after correcting for the
clustering effect, and the blue solid line after correcting by the background. The dashed
black, blue and red lines show the sample without being corrected and the expected level of
contamination due to  background and the clustering, respectively. The numbers above each bin
are the number of observed satellites within each mass interval. The vertical dashed area
corresponds to the mass ratio region where incompleteness in the satellite samples could play
a role. The upper panel is for the spectroscopic sample and the bottom panel the results
obtained using the photometric sample.} 

\label{merging_channel} 
\end{figure}

In Fig. \ref{merging_channel} and Table \ref{merging_channel_tab}, we show the most likely
merging channel of our massive ellipticals.  We show our results after being corrected by the
effect of the background and clustering. We have assumed Poissonian errors to estimate our
error bars. The total amount of mass contained in the satellite population down to 1:10
compared to the total amount of mass in the hosts is $\sim$6 per cent (using the spectroscopic
sample) or $\sim$5.5 per cent (using the photometric sample). Down to 1:400 this value  is $\sim$8 per cent
of the total amount of mass contained in their hosts. The largest contributor to the
satellite mass are those satellites with a mass ratio from 1:2 to 1:5 ($\sim$28 per cent of the
total mass of the satellites in the photometric sample). If the satellites eventually infall
into the host galaxies, the merger channel will be largely dominated by satellites with a
mass ratio below 1:10 (which have 68 per cent of the total mass in satellites). This is again a
result of the approximately constant number of satellites we find when we move towards less
and less massive satellites. For this reason, the decrease of the stellar mass in the
satellites can not be compensated by a larger number of these objects and the main driver of
mass accretion is provided by the larger satellites. If the theoretical expectation holds and
the merger time-scales are shorter for the most massive satellites, the mass growth due to the
larger satellites will be even more important than the result show in Fig.
\ref{merging_channel}. We discuss the consequence of this result in the next section.

\begin{table*}

\caption{The merging channel of present-day massive ellipticals. The table shows the
contribution of the satellite mass enclosed in each satellite mass bin to the total mass
confined in their hosts. We show the observed fraction (in per cent) and the fractions (in per cent) after
correcting of background and clustering.}

\begin{tabular}{lccc}
\hline\hline
$M_{\rm Sat}/M_{\rm Host}$ & $\Psi$ (Obs) &  $\Psi$ (Sim) & $\Psi$ (Clu) \\
\hline
                    &   &      Spectroscopic sample      & \\
\hline
0.5-1.0&$     1.67\pm     0.33$&$     1.67\pm     0.32$&$     1.40\pm     0.31$\\
0.2-0.5&$     2.68\pm     0.27$&$     2.64\pm     0.27$&$     2.08\pm     0.24$\\
0.1-0.2&$     2.39\pm     0.19$&$     2.37\pm     0.18$&$     2.07\pm     0.17$\\
\hline
Total&$     6.77\pm     0.47$&$     6.68\pm     0.46$&$     5.55\pm     0.42$\\

\hline
                 &   &        Photometric sample     & \\
\hline
0.5000-1.000&$     2.42\pm     0.35$&$     2.02\pm     0.33$&$     1.64\pm     0.29$\\
0.2000-0.500&$     3.14\pm     0.27$&$     2.46\pm     0.24$&$     2.00\pm     0.21$\\
0.1000-0.200&$     1.97\pm     0.15$&$     1.50\pm     0.13$&$     1.26\pm     0.12$\\
0.0500-0.100&$     1.51\pm     0.09$&$     1.15\pm     0.08$&$     0.93\pm     0.07$\\
0.0200-0.050&$     1.30\pm     0.06$&$     0.95\pm     0.05$&$     0.82\pm     0.05$\\
0.0100-0.020&$     0.62\pm     0.03$&$     0.37\pm     0.02$&$     0.34\pm     0.02$\\
0.0050-0.010&$     0.41\pm     0.02$&$     0.20\pm     0.01$&$     0.17\pm     0.01$\\
0.0025-0.005&$     0.24\pm     0.01$&$     0.09\pm     0.01$&$     0.08\pm     0.01$\\
\hline
Total&$    11.59\pm     0.48$&$     8.75\pm     0.44$&$     7.21\pm     0.39$\\
\hline\hline

\end{tabular}
\label{merging_channel_tab}
\end{table*}

\section{Discussion}\label{sec:discussion}

The results presented in this paper can be used as a z$\sim$0 reference for the study of the evolution
of satellites around massive elliptical galaxies with cosmic time. As we mentioned in the Introduction,
several studies have addressed recently the evolution of the number of satellites of massive ellipticals
with redshift both observationally and theoretically. These previous studies have shown that the
fraction of massive elliptical galaxies with satellites of a given mass ratio has remained constant
since at least z$\sim$2. However, those works disagree on the number of massive galaxies having
satellites, being more abundant the satellites in the simulations than in the observations. In this
paper, which is less affected by incompleteness at small stellar masses than previous works, we can
readdress this question  and see how the theoretical expectations compare to the observational data in
the nearby Universe.

\subsection{Comparison with theoretical expectations}

\citet{Quilis2012} have estimated, using the Millennium simulation, what is the expected
fraction of massive galaxies having satellites with a mass ratio down to 1:10 and down to
1:100 within a sphere of R=100 kpc. They have conducted such studies exploring galaxies from
z=2 to now using three different semianalytical models. At z=0, the theoretical expectations
suggest that the fraction of massive galaxies having satellites with mass ratio down to 1:10
ranges from 0.3 to 0.4. Observationally, we have found 0.23 after the background correction
and 0.20 after the clustering correction (using the spectroscopic sample).  These values are
lower than the theoretical expectations. If we focus our attention now towards satellites
with lower masses, we find that the theoretical expected fraction of massive galaxies with at
least a satellite with a mass ratio down to 1:100 is between 0.6 and 0.7. These values are
again larger than the observed data, where we find 0.52 after the background correction and
0.45 after the clustering correction (using the photometric sample). It is worth stressing
that this discrepancy among the theoretical and the observational results can not be
explained due to the different volumes of exploration used in both works: a spatial sphere of
100 kpc in \citet{Quilis2012} and a cylinder in redshift here. Because of the way we have
selected our galaxies in redshift (using up to 3$\sigma$ away from the redshift of the galaxy
host), basically all the satellite galaxies in the line of sight of the host within a
projected radial distance of 100 kpc are taken. In that sense, at comparing with the
Millenium simulation our number of observed satellites should be an upper limit (as they are
only restricted to 100 kpc in depth). As the number of theoretical satellites is larger than
observed, we can confidently claim that there is a discrepancy with our observations. Also
note that the inner hole in the number of observed satellites produced by the presence of the
galaxy host (see Fig. \ref{splitinmass}) only outshone a small fraction of the probed area
($\sim$4 per cent for the photometric sample). Consequently, this cannot help to explain the
discrepancy either. Finally, a potential lost of  satellites in the work by
\citet{Quilis2012} due to resolution effects will also increase the discrepancy between the
simulations and the observations.

We think that the excess of satellites found in the Millenium simulation is due to the
inability of the semi-analytical models to reproduce \citep[see the Fig. 20 of][]{Guo2011} the
right clustering of low-mass (M$_\star$$<$6$\times$10$^{10}$M$_\odot$) galaxies at small
scales (R$<$1 Mpc). The model substantially overproduce the clustering at small scales for
these low-mass objects. \citet[][]{Guo2011} suggest that this discrepancy could be due to the
large $\sigma_8$=0.9 adopted in the simulation compared to the observational most likely
value of $\sigma_8$=0.834$\pm$0.027 \citep[][]{Planck2013}.

Another quantity that we can compare with the theoretical predictions is the average projected radial
distance of the satellites to their galaxy hosts. \citet{Quilis2012} found that the average projected
radial distance  ranges within 35-45 kpc. Here we find a value which is $\sim$59 kpc and that could
be decreased down to 49-51 kpc if the incompleteness of satellites in the central region of search is
accounted for. It is worth noting that in our rough correction of the incompleteness, we have assumed
an equal probability with the radius of the satellite distribution. In practice, it is likely that
there would be an excess of probability of finding more satellites closer to the galaxy hosts than what
we have assumed. If this were the case, our estimation of a projected (corrected) radial distance of
$\sim$50 kpc would be an upper limit. Moreover, we warn the reader that a direct comparison among the
theoretical and observational results is not straightforward.  \citet{Quilis2012} search for satellites
within a sphere of R=100 kpc whereas the observational search actually resembles a cylinder. In this
sense, the observed radial distance can be again more prone towards larger radial distances than the
theoretical study. Summarizing, the average radial distance of the satellites in the simulation and in
the observations, once the incompleteness in the observations is accounted for, could be in reasonable
agreement.

\subsection{Comparison with previous data}

We can now draw our attention to the fraction of massive galaxies with satellites found in other works
at z=0. \citet{Liu2011}, exploring MW-like galaxies (i.e. objects which are less massive than our host
galaxies), found that only 14 per cent of those objects have at least a satellite within R=100 kpc down to
1:10 mass ratio (private communication). In Fig. \ref{alldata} we have compared this observation with our
results. Liu's number is slightly below our findings (20-23 per cent in the spectroscopic sample) but we think
it is as expected taking into account that the number of satellites depends on the mass and the color
of the host \citep{Wang2012}. Larger the mass of the host larger is the number of its satellites.

A more direct comparison with our range of mass for the host galaxies can be done with the results of
\citet{Wang2012}. In that work, the authors have not segregated the galaxies using visual morphology but
colors. We will compare our results with the ones found for their red hosts, likely the most similar to our
elliptical galaxies. Another important difference is that they have done the search of satellites within a
radius of 300 kpc.  Consequently, to allow a direct confrontation with this data set we have repeated our
analysis using such radius and only for those host galaxies with 11.1$<$logM$_\star$$<$11.4 (their
green line in their Fig. 7). We get the following results: (a) for satellites with logM$_\star$=10, $N_{\rm
Sat}/N_{\rm Host}/\log M_\star$=3.2-4.5 (ours) and 3-4 (Wang \& White) and (b) for satellites with
logM$_\star$=9, $N_{\rm Sat}/N_{\rm Host}/\log M_\star$=4.5-6.1 (ours) and 6-7 (Wang \& White).
This agreement is remarkable taking into account the different techniques and selection of the galaxy hosts.

Finally, we can make a comparison with \citet{Marmol-Queralto2012}. These authors conducted a
similar analysis to what we have done here but for galaxies at higher redshifts
(0.2$<$z$<$2). We  compare our numbers (see Fig. \ref{alldata}) with the galaxies they
classified as ellipticals in their lower redshift range (0.2$<$z$<$0.75). They found that the
fraction of massive galaxies with satellites down to 1:10 are 0.23-0.28. Here we find
0.20-0.24 (in the spectroscopic sample depending whether the clustering or background
correction has been applied) and 0.16-0.19 (in the photometric sample), which is slightly
below but in reasonable agreement with those results.  Summarizing, our results seem to agree
pretty well with previous studies in those ranges of masses where a direct comparison can be
conducted.

\subsection{The merging channel in the present-day Universe}

There is currently a strong debate about what is the favoured merging channel which explains the
dramatic increase in size of the massive galaxies in the last 11 Gyr. Whereas there is a growing
agreement \citep[see e.g.][]{Trujillo2011} that the size evolution can not be entirely explained by
internal mechanisms like AGN activity \citep[][]{Fan2008,Fan2010,Ragone2011}, it is not clear what is
the relevance of major versus minor merging in the growth of the galaxies. On one hand, major mergers
\citep[e.g.][]{Ciotti2001,Nipoti2003,Boylan-Kolchin2006,Naab2007} seem to be very scarce \citep[at
least since z$\sim$1;][]{Bundy2009,deRavel2009,Wild2009,Lopez-Sanjuan2010,Kaviraj2011}  to play a major role in the
growth of the galaxies. On the other hand, minor merging \citep[favoured theoretically for its
efficiency on increasing the size of the
galaxies;][]{Khochfar-Burkert2006,Maller2006,Hopkins2009b,Naab2009}  confronts some problems with the
number of satellites found at z$\sim$1 \citep[e.g.][]{Ferreras2014}.

\citet{Oser2012} have suggested that the most likely scenario is an increase in size and mass of the
massive galaxies through satellites having a mass ratio of 1:5. The results of our paper cannot solve
this question directly, as we would need to explore the satellite distribution at different cosmic times
to address this issue. However, motivated by the observed \citep{Marmol-Queralto2012} and theoretically
expected \citep{Quilis2012} constancy of the fraction of massive galaxies having satellites since z=2, we
can speculate about  the merging channel of massive ellipticals back in time. 

If galaxies at high-z would follow the same mass distribution than the one observed in the nearby
Universe, then the observations will suggest that the mass and size increase of the massive elliptical
galaxies will be dominated by satellites with  mass ratio within 1:2 to 1:5 \citep[see
also][]{Lopez-Sanjuan2012}. Note, however, that this statement assumes that the merger time-scale is
independent on the mass ratio between the satellites and the host galaxies. More realistic scenarios
\citep[e.g.][]{Jiang2013} suggest that the merger time-scale rises as the mass ratio between both
galaxies increases. In this sense, the smaller satellites will take significantly more time to merge
with the host galaxies than the bigger satellites. Based on this, what we can claim with some
confidence is that  low-mass satellites with mass ratio below 1:10 would play a minor role in the mass
increase of the host galaxies. They would be just very small in number to contribute to the mass growth
plus they will have very large time-scales to efficiently infall into the massive ellipticals.

\section{Summary and Conclusions}\label{sec:conclusions}

In this paper we have explored the properties of the satellite population around massive,
visually classified, ellipticals in the nearby Universe (z$<$0.1). Our aim has been to
robustly quantify the distribution of satellites around this type of objects to create a
local reference for future studies of the evolution of the satellites with cosmic time. We
have based our analysis on a sample of $\sim$1000 massive ellipticals obtained from the
catalogue of NA10. To explore the satellite distribution around these objects we
used the spectroscopic NYU VAGC and the photometric photo-z SDSS DR7
catalogue. This has allowed us to explore satellites down to a mass ratio of 1:400.

Our satellite galaxies have been identified within a projected radial distance of 100 kpc around the
hosts. A careful analysis of the background and clustering contamination has been done. We have found
that only 20-23 per cent of the massive ellipticals has at least one satellite with a mass ratio down to 1:10.
This number increases to 45-52 per cent if we explore satellites down to 1:100 and to $>$60-70 per cent if we go down
to 1:400. The average observed projected radial distance of the satellites to the host is $\sim$59 kpc
(which can be decreased down to, at least, $\sim$50 kpc if we account for incompleteness effects). The
observed number of satellites are lower than the predictions from theoretical expectations \citep[see
e.g.][]{Quilis2012}. It will be worth exploring in the future whether this disagreement with the
theoretical models increases for satellites with lower masses (i.e. M$_\star$$\lesssim$10$^{8}$M$_\odot$).

Finally, the number of satellites per galaxy host only increases very mildly at decreasing the
satellite mass. The fraction of mass which is contained in the satellites down to a mass ratio of 1:400
is $\sim$8 per cent of the total mass contained by the hosts. The largest contributor to the  mass enclosed by
the satellites are those satellites with a mass ratio from 1:2 to 1:5 (in fact, $\sim$28 per cent of the total
mass of the satellites is within objects with such mass ratio). If the satellites eventually infall
into the host galaxies, the merger channel will be largely dominated by satellites with a mass ratio
down to 1:10 (these satellites have 68 per cent of the total mass in satellites). Futures studies \citep[see,
for instance,][]{Ferreras2014}, exploring how is the mass distributed among the satellites of higher
redshift objects, will allow to test whether the currently suggested most likely scenario for
explaining the increase in size and mass of the massive galaxies \citep[i.e. by the infall of
satellites having a mass ratio of around 1:5;][]{Oser2012} is favoured or not.

\section*{Acknowledgments}

We thank the referee for constructive comments and careful reading of our manuscript.
We are grateful to Juan Betancort for his help on the statistical analysis conducted in this
paper. Vicent Quilis and Thorsten Naab help to understand better the comparison between the
theoretical and observational results. Carlos L\'opez Sanjuan did some interesting
observations on the background corrections applied to the data. This work has been supported
by the ``Programa Nacional de Astronom\'{\i}a y Astrof\'{\i}sica'' of the Spanish Ministry of
Science and Innovation under grant AYA2010-21322-C03-02. EMQ acknowledges the support of the
European Research Council via the award of a Consolidator Grant (PI McLure). This project has
made use of data from the Sloan Digital Sky Survey (SDSS). Funding for the SDSS has been
provided by the Alfred P. Sloan Foundation, the Participating Institutions, the National
Science Foundation, the U.S. Department of Energy, the National Aeronautics and Space
Administration, the Japanese Monbukagakusho, the Max Planck Society, and the Higher Education
Funding Council for England. The SDSS Web Site is \url{http://www.sdss.org/}.

The SDSS is managed by the Astrophysical Research Consortium for the
Participating Institutions. The Participating Institutions are the American
Museum of Natural History, Astrophysical Institute Potsdam, University of
Basel, University of Cambridge, Case Western Reserve University, University of
Chicago, Drexel University, Fermilab, the Institute for Advanced Study, the
Japan Participation Group, Johns Hopkins University, the Joint Institute for
Nuclear Astrophysics, the Kavli Institute for Particle Astrophysics and
Cosmology, the Korean Scientist Group, the Chinese Academy of Sciences
(LAMOST), Los Alamos National Laboratory, the Max-Planck-Institute for
Astronomy (MPIA), the Max-Planck-Institute for Astrophysics (MPA), New Mexico
State University, Ohio State University, University of Pittsburgh, University
of Portsmouth, Princeton University, the United States Naval Observatory, and
the University of Washington.

\bibliography{sgmeg_MNRAS_matching}
\bibliographystyle{mn2e}

\end{document}